\documentstyle[floats,twocolumn,aps,prl]{revtex}

\begin{document}
\draft
\title{
\vspace*{-1cm}\hfill
 {\tt }
       \vspace{1cm}\\
Anisotropy of Growth of the Close-Packed Surfaces of Silver
}
\author{Byung Deok Yu and Matthias Scheffler} 
\address{Fritz-Haber-Institut der Max-Planck-Gesellschaft,
Faradayweg 4-6, D-14195 Berlin-Dahlem, Germany}

\date{\today}

\twocolumn[
\maketitle


\vspace*{-10pt}
\vspace*{-0.7cm}
\begin{quote}
\parbox{16cm}{\small

The growth morphology of clean silver exhibits a profound anisotropy: The
growing surface of Ag\,(111) is typically very rough while that of Ag\,(100)
is smooth and flat. This serious and important difference is unexpected,
not understood, and hitherto not observed for any other metal. Using
density functional theory calculations of self-diffusion on flat and
stepped Ag\,(100) we find, for example,  that at flat regions a hopping mechanism is
favored, while across step edges diffusion proceeds by an exchange process.
The calculated microscopic parameters  explain the experimentally reported growth properties.

} 
\end{quote} ]

\narrowtext

%
The morphology and microscopic and mesoscopic quality of surfaces
is one of the major concerns in 
investigations of epitaxial growth. Often the growth mode and the resulting
surface quality can be altered by varying the growth conditions such as the
substrate temperature and deposition rate, or by introducing defects or
surfactants (see Ref. \cite{Scheff96} and references therein). Interestingly,
different substrate orientations sometimes exhibit a noticeably different
behavior. Silver is probably one of the extreme examples: Homoepitaxy on clean
Ag\,(111) gives rise to very rough surfaces while on clean Ag\,(100) the
growing surface remains smooth and flat. This result has been found by
experimental studies using reflection high-energy electron diffraction
(RHEED) intensity oscillations~\cite{Suzuki} showing for Ag\,(100) at
temperatures between 200 and 480 K a smooth two-dimensional (2-D) growth
behavior. In contrast, for Ag\,(111)  RHEED intensity oscillations are absent
at all temperatures, suggesting that there is no 2-D growth at all. This
finding for Ag\,(111) was confirmed by x-ray reflectivity experiments~\cite{Vegt}
and scanning tunneling microscopy~\cite{Vrijmoeth,Bromann}.
The latter studies exposed that the surface becomes deeply fissured
with ``mountains'' as high as  30-40 atomic layers.
The analysis gave that silver adatoms on top of existing islands encounter a
step-edge barrier hindering their descent. This barrier is noticeably
higher (by 0.15~eV~\cite{Vrijmoeth}) than the diffusion barrier on the flat
terrace which
explains why deposited atoms which land on islands will stay on the island
giving rise to the observed multi layer growth of silver in the (111)
orientation.

In general, the existence of a step-edge barrier~\cite{Schwoebel} is plausible because 
of the low coordination of the step-edge atoms~\cite{Stumpf}.
Thus, a priori a significant difference in growth mode between the two similarly
close packed surfaces of fcc metals is not to be expected, and in fact, has
not been reported so far for other systems.
To date various semi-empirical embedded-atom methods have been 
used to study fcc metals.
However, we note that such semi-empirical treatments may not 
be predictive, although it is 
often assumed that such approaches may explain general trends. 
Thus, a more elaborate and reliable theory is 
necessary to find out if the growth anisotropy of  silver
is really special and what actuates the smooth growth of
the (100) surface. Below we will answer these questions.

We performed density functional
theory studies. Most calculations were done with the exchange-correlation
functional treated in the local-density approximation (LDA); at  
the important geometries we repeated the calculations
using the generalized gradient approximation (GGA)~\cite{Perdew}.
We used an updated version of the computer code described in 
Ref.~\cite{Stumpf2} together with norm-conserving, fully separable
pseudopotentials~\cite{Troullier,Gonze}. The surface was simulated by a repeating slab
in which typically three atomic layers and a 10.35 \AA{} vacuum region are
included. In the lateral directions we took a $(3 \times 3)$ periodicity, 
which we tested gives that the artificial (unwanted) 
adatom-adatom interaction is sufficiently weak and in fact negligible for
the questions of concern. The LDA (GGA) pseudopotentials were created
using the nonrelativistic (relativistic) scheme of Troullier and 
Martins~\cite{Troullier} and
Kleinman and Bylander~\cite{Gonze}, as described by 
Fuchs {\em et al.}~\cite{Fuchs}. The details of the potential,
such as presentation of its logarithmic derivatives and
electronic hardness properties will be published elsewhere~\cite{Yu2}.
The 4$d$ states are included as valence states and the basis-set consists of
plane waves up to a kinetic energy of 40 Rydberg.
For the silver bulk this treatment gives a lattice constant
$a_0=$ 4.14 (4.18)~\AA\, and a bulk modulus of $B_{0}=$ 0.99 (0.90)~Mbar.
These are the LDA results and the values in brackets are
obtained using the GGA. 
In these values the zero point vibrations are not taken into account.
The agreement with $T \rightarrow 0$~K experimental data
is good ($a_0^{\rm exp} = 4.07$~\AA\, and a bulk modulus of $B_{0}^{\rm exp}=$
1.02~Mbar), as is that with other calculations~\cite{Meth92b,Khein}.
For the {\bf k} summations in the surface calculations we took nine equidistant
{\bf k} points in the surface Brillouin zone (SBZ) of 
the $(3 \times 3)$ cell (avoiding $\Gamma$ point).
We relaxed the adatoms and the top-layer atoms,
keeping the other Ag atoms in their theoretical-bulk positions.
All geometries were optimized until the  remaining forces were smaller than
0.05~eV/\AA.
In order to attain fast convergence of the iterative solution of the 
Schr{\"o}dinger equation we found it important to start with initial
wave functions obtained from a mixed basis set of pseudo atomic orbitals and 
plane waves with a 4 Ry cutoff, as developed
by Kley {\em et al.}~\cite{Kley}.
The number of substrate layers is admittedly small, we 
therefore note that the adatom is adsorbed only on one side of the slab
(see also Ref.~\cite{Neugebau92}).
\begin{figure}[b]
  \leavevmode
  \includegraphics{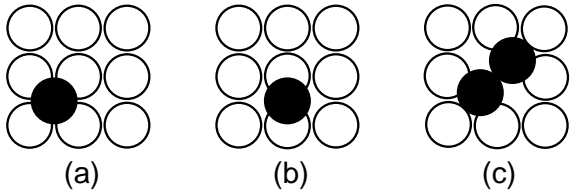}
  \vspace*{3.1cm}
\caption{Schematic top view of geometries at (a) the fourfold hollow site,
 (b) the transition state for the hopping diffusion (the twofold bridge site), 
 and (c) the transition state for the exchange diffusion. 
 The solid and open circles represent the ad- and surface atoms, respectively.
         }
\label{geo}
\end{figure}

We first summarize our LDA (GGA) results for the clean Ag(100) surface. 
The top-layer relaxation is $\Delta d_{12}= -2.0$\%{}$d_0$ ($-1.4$\%{}$d_0$),
where $d_0$ is the interlayer spacing in the bulk. The surface energy per atom is 
$\sigma =$ 0.61~eV (0.48~eV), and the work function is $\phi =$ 4.38~eV (4.30~eV).
The agreement with previous full-potential 
linear-muffin-tin-orbital (FP-LMTO) calculations~\cite{Meth92b} 
(only LDA results exist so far) is very good.
We now address the energetics of adsorption and 
diffusion of a Ag adatom on the flat  Ag\,(100) surface. 
We find that the stable adsorption site for a Ag adatom is the fourfold hollow
(Fig.~\ref{geo}(a)), i.e. the highest coordination site, 
as would be expected for a noble metal.
The four nearest neighbors of the adatom distort laterally, opening
the hollow site even further.
The bond length between the adatom and its neighbors is
2.78~\AA{}, i.e., 5~\%{} shorter than the interatomic distance in the
bulk. This follows the typical trend, namely that bond strength
per bond decreases with coordination and correspondingly, bond length
increases with coordination.
Diffusion  on the (100) surface of fcc metal may proceed  by a hopping or an
exchange process. For the hopping process the transition state is
the twofold coordinated bridge geometry (see Fig.~\ref{geo}(b)).   
For the exchange process the transition state consists of
two  atoms sitting over a surface vacancy (see (Fig.~\ref{geo}(c)).
The exchange process has been theoretically predicted by Feibelman~\cite{Feibelman} 
to be active at Al\,(100) and stabilized by a covalent like bonding in the transition
state geometry. Experimental evidences for diffusion by exchange
have been seen for Pt\,(100)~\cite{Kellogg} and Ir\,(100)~\cite{Chen}. 
For Cu\,(100) Hansen {\em et al.}~\cite{Hansen}, using the effective medium approach,
predicted exchange diffusion to have lower energy barriers than a
hopping.
\narrowtext

\begin{table}[b]
\caption{Convergence tests for the LDA diffusion barriers $E_d$ on flat
 surfaces with respect to the cutoff energy $E_{cut}$,
 the surface cell size, 
 the number of atomic layers $N_l$, and the number of {\bf k} points in the SBZ
 $N_{k}$. }
\begin{tabular} {cccccc} 
$E_{cut}$ & surface cell & $N_l$ & $N_k$ & \multicolumn{2}{c}{$E_d$ (eV)} \\
(Ry)             &   size        &       &       & \multicolumn{1}{c}{hopping} &
                                               \multicolumn{1}{c}{exchange} \\
\tableline
40 & 3$\times$3 & 3 & 9 & 0.52 & 0.93 \\
\tableline
50 & 3$\times$3 & 3 & 9 & 0.52 & 0.94 \\
40 & 2$\times$2 & 3 & 16 & 0.50 & 0.95 \\
40 & 2$\times$2 & 4 & 16 & 0.51 & 0.95 \\
40 & 3$\times$3 & 4 & 9 & 0.53 & 0.92 \\
40 & 3$\times$3 & 3 & 16 & 0.51 & 0.96 \\
\end{tabular}
\label{test}
\end{table}

For Ag\,(100) we find that the exchange process is clearly unfavorable.
The energy barriers for hopping diffusion are
0.52~eV (LDA) and  0.45~eV (GGA) which is close to the results
of the FP-LMTO calculation~\cite{Boisvert},
which got 0.50~eV (LDA). The
semi-empirical study of Liu {\em et al.}~\cite{Liu} obtained
with the embedded-atom method (EAM) a similar values, namely 0.48~eV.
The difference between the LDA and the GGA result is small
(0.07~eV).  
In the bridge site the bond length between the adatom and its two 
neighbors is 2.69 \AA{}, i.e. 3 \%{} shorter than in the fourfold hollow.
This again follows the well known trend:
Each of the two bonds at the bridge site is stronger than each of the
four bonds at the hollow site.

For the exchange diffusion we obtain an energy  barrier
of 0.93~eV (LDA) and 0.73~eV (GGA) which is much higher than
that for a hopping process. Thus, we can rule out that 
the exchange process will play
a role for self-diffusion at flat regions of Ag\,(100).
It is interesting that for the exchange geometry the difference
between LDA and GGA is noticeable (0.20~eV). This is somewhat plausible, 
because in the transition state of the exchange process the bonds are more
localized.

Some results of convergence tests for the diffusion barrier on flat surfaces are 
given in Table~\ref{test}. The tests have been done by additional calculations
using a more extensive set of parameters, namely a (2$\times$2) cell,
4-layer slab, basis with 50 Ry energy cutoff, and 16 {\bf k} points
in the SBZ. Increasing the cutoff energy from 40 Ry to 50 Ry changed the
diffusion barriers by less than 0.01 eV and increasing the number of atomic
layers from 3 to 4 results in changes of only 0.01 eV in the barriers.
Increasing the number of {\bf k} points from 9 to 16 changed the results by
less than 0.03 eV. Our calculations suggest that the results for the calculated
energy barriers and other total-energy differences are accurate to 0.05 eV.
Thus, for the present study our numerical accuracy is 
sufficiently high.

\begin{figure}[b]
  \leavevmode
  \includegraphics{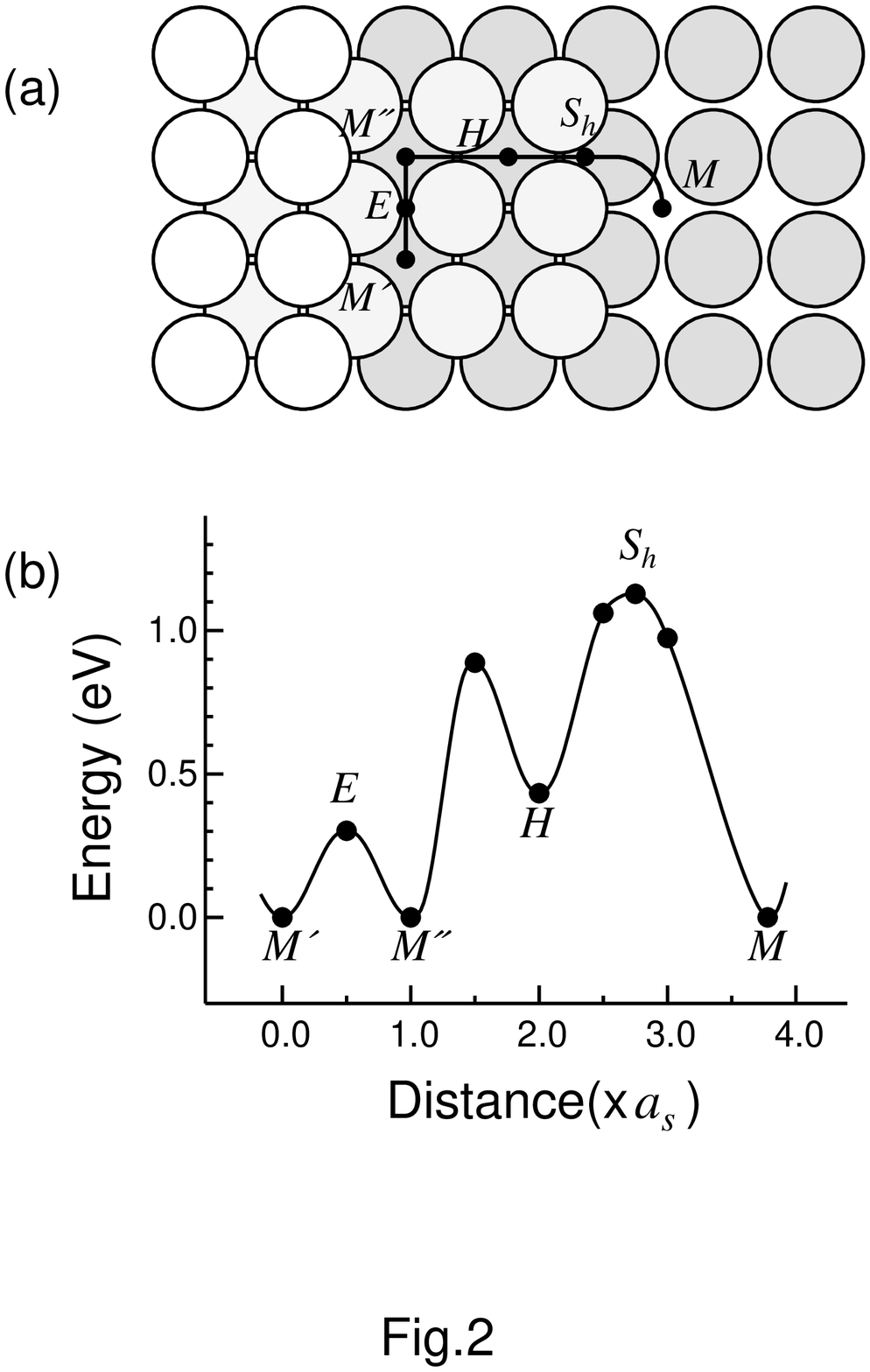}
  \vspace*{9.5cm}
\caption{Total energy of an Ag adatom diffusing along the 
         indicated path by a hopping process, 
         calculated within the LDA.
         A top view of the vicinal (511) surface is shown in (a).
         The step edges are aligned along the [0\={1}1] direction.
         The distance is given in unit of the surface lattice constant 
         $a_s=2.92$ \AA.
         }
\label{hopping}
\end{figure} 
While at the flat regions  exchange diffusion is found to be unimportant
the situation at steps is somewhat different.
We considered a (511) surface which is vicinal to (100).
This surface consists of (100) terraces which are 3 atoms wide.
The step edges are perpendicular to the [100] and the [011] directions
(see Fig.~\ref{hopping}(a)).
The periodicity along the step edge is taken to be
three surface lattice constants.
Figure~\ref{hopping}(b) displays the results obtained for the hopping
process. There are two stable sites: The hollow site ($M$) at which
the adatom is fivefold coordinated and
the hollow site ($H$) on the terrace, at which the adatom coordination is four.

\begin{figure}[b]
  \leavevmode
  \includegraphics{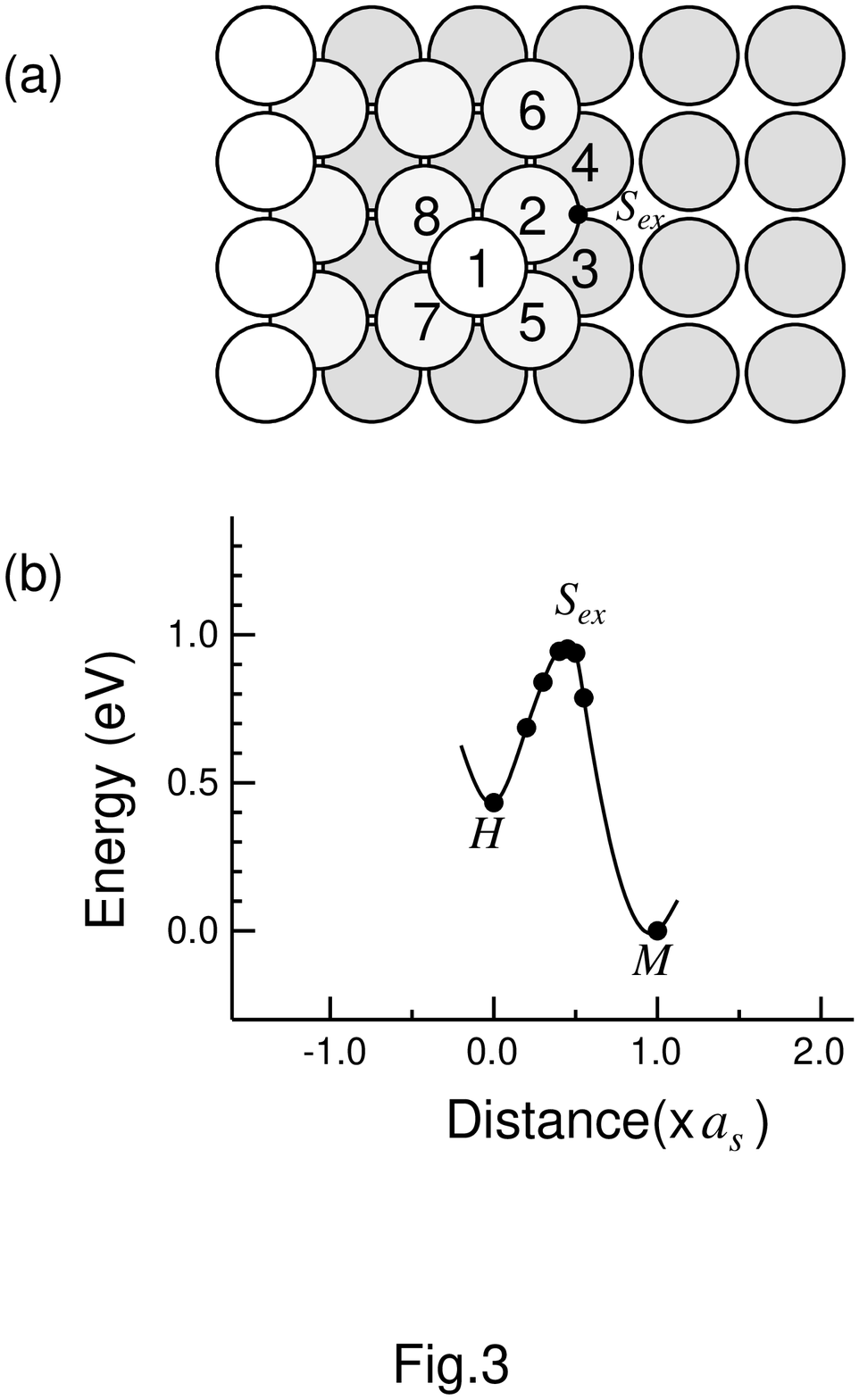}
  \vspace*{9.5cm}
\caption{Total energy of an Ag adatom diffusing across a step by an
         exchange process, 
         calculated within the LDA.
         In (a) adatom 1 is adsorbed at site $H$.
         The total energy as a function of  
         the distance of the step-edge atom 2 from the undistorted step edge
         is shown in (b). 
         }
\label{exchange}
\end{figure}
It is plausible that at site $M$ the adatom is bound best.
The energy difference between the $M$ and $H$ sites is 0.43~eV (LDA)
and  0.32~eV (GGA). 
The energy at the $H$ site is in fact the same (within our accuracy of 0.05 eV)
as that at flat regions of the surface. This result is found by comparing
our total energies of the flat and the stepped surfaces and by doing additional
calculations for a (711) surface, which has longer terraces. 
The ``additional energy barrier'' for rolling over the ledge from $H$ to $M$
is $\Delta E_{\rm step}^{\rm Ag\,(100)} = 0.18$~eV (LDA) and 0.10~eV (GGA). 
Thus, it is by about
30~\% (20~\%) higher in the LDA (GGA) than the diffusion barrier at the flat surface.
The transition state for the roll over process is identified to
be near the bridge site at the ledge ($S_h$ in Fig.~\ref{hopping}(a)).
The other possibly to reach the geometry $M$ from $H$ 
is via an exchange, where atom 1 replaces atom 2 and subsequently 
the latter moves to the fivefold coordinated site $M$ (see Fig.~\ref{exchange}).
Within our numerical accuracy for energy differences ($\approx 0.05$~eV)
the calculated energy barrier of 0.52~eV (LDA) and 0.45~eV (GGA) is almost identical 
to that of the hopping diffusion at the  terrace.
At the transition state geometry the atom 2 is located near the bridge site
formed by two step-edge atoms 3 and 4 on the lower terrace
($S_{ex}$ in Fig.~\ref{exchange}(a)).
The interesting feature is that the step-edge atom 2 is situated closer
to atom 4 to form a bond with atom 6. Each of the two top-most atoms 1 and 2
has five bonds with neighbors, 
while at the saddle point of the exchange diffusion on the flat surface 
each of the two top-most atoms has only four bonds (Fig.~\ref{geo}(c)).
The origin of the lower diffusion barrier of the adatom at the edge 
is thus the additional bonds formed at the obtained saddle point.  
We have indeed found that {\it there is no additional energy barrier to
descend from an upper to a lower terrace}.
This finding provides a natural explanation for the
smooth 2-D growth observed in homoexpitaxy on the Ag\,(100) surface.

Our total energy calculations clearly show that the step-down diffusion
proceeds by an exchange process. 
Inspection of the geometry (see Fig.~\ref{exchange}) makes this theoretical finding
very plausible.
For this  process atom 2 in Fig.~\ref{exchange} has to move only a short distance,
namely to the next hollow site, passing over a bridge position; this
displacement is similar to the hopping at the flat surface. Because
atom 1 follows in close contact to atom 2, the local coordination of
atoms 1, 2, 5, and 8 always remains high, making the process
energetically favorable. In contrast, the roll over process (see
Fig.~\ref{hopping}) is more involved, i.e. the diffusing atom has to proceed
a much longer distance on which it even has to pass a nearly-on-top
site, when  turning from the $S_h$ point towards $M$.

We note in passing that exchange processes will also play a role 
at steps at fcc\,(111) surfaces but here the displacements are more
complex and the additional step edge barrier is not diminished.
We also note that at aluminum surfaces, the system studied best so far,
the exchange process is supported by the formation of covalent bonds,
which is made possible by the  $sp$ valence electrons~\cite{Feibelman}. 
For silver we find that the electron density does not reflect a pronounced 
covalency effect, which is indeed the expected behavior for a noble metal. 
This result explains why the exchange process does not happen at the flat
Ag\,(100) surface and has only a minor effect at the close-packed
steps at Ag\,(111).

Finally we  comment on the shape of islands.
Key parameters that determine the shape of islands are the types of steps and
the mobility of atoms along step edges. Since there are two different types of
steps
for fcc\,(111), in thermodynamic equilibrium islands have a hexagonal shape.
For fcc\,(100) we find only one type of step as shown in Fig.~\ref{hopping}(a) 
(the other step has a \{110\} microfacet and needs much higher energy to form).
Thus, the expected equilibrium shape of islands is a square.
The diffusion barrier along step edges determines the roughness of steps. 
We find that adatom diffuses along step edge by a hopping process.
The barrier is significantly
lower than the surface diffusion barrier $E_d$. 
This lower energy barrier (LDA: 0.30~eV; GGA: 0.27~eV) indicates that 
Ag islands formed on Ag\,(100) should be 
compact. Atoms which reach the step edges will certainly be able to diffuse parallel
to the steps and thus local thermal equilibrium is attained. 
We therefore expect rather straight step edges and 
no fractally shaped islands. 
The lower barrier for diffusion parallel to steps can be understood 
in terms of variation in coordination 
from five to four as
the adatom moves from site $M^{\prime}$ to site $E$ (Fig.~\ref{hopping}(a)), 
while for diffusion on flat terraces the coordination
varies from four to two.

In summary, we presented density functional theory
calculations for various microscopic diffusion processes at Ag\,(100).
Ag adatoms are found to diffuse across flat terraces by a hopping process.
Adatoms approaching step edges descend from the upper to the lower terrace
by an exchange process and the obtained energy barrier is found to be 0.45~eV
in the GGA,
almost identical to the barrier at flat regions. 
Thus, there is  no additional energy
barrier to diffuse across step edges, in sharp contrast to the self-diffusion 
at Ag\,(111).
This implies good inter-layer mass transport for deposition at and growth
of silver (100) giving rise to a smooth surface. In contrast, the
additional step-edge barrier which exists at the silver (111) surface (and
typically at other surfaces as well) gives rise to the rough growth of Ag\,(111).
Also our finding that the step down motion proceeds by exchange is expected to
apply for other noble metals and their left neighbors (but this isn't guarantee that
the additional step-edge barrier vanishes completely). 

We thank A.~Kley, S.~Narashimhan, G.~Boisvert, C.~Ratsch, and P.~Ruggerone
for helpful discussions.
B.~D.~Yu gratefully acknowledges a fellowship from the Alexander von Humboldt
Foundation. 



\begin{references}
\bibitem{Scheff96} M. Scheffler, V. Fiorentini,
 and S. Oppo, In: Proc. 2nd German-Australian Workshop on Surface Science,
January 24-28, 1994, Eds. R. MacDonald, E. Taglauer, K. Wandelt.
Springer-Proc. in Physics (SPP), Springer, Berlin 1996.
\bibitem{Suzuki} Y. Suzuki, H. Kikuchi, and N. Koshizuka, Jpn. J. Appl. Phys.
  {\bf 27}, L1175 (1988). 
\bibitem{Vegt} H. A. van der Vegt, H.M. van Pinxteren, M. Lohmeier, E. Vlieg, and
 J.M.C. Thornton, Phys. Rev. Lett. {\bf 68}, 3335 (1992).
\bibitem{Vrijmoeth} J. Vrijmoeth, H.A. van der Vegt, J.A. Meyer, E. Vlieg, 
and R. J. Behm, Phys. Rev. Lett. {\bf 72}, 3843 (1994).
\bibitem{Bromann} K. Bromann, H. Brune, H. R\"oder, and K. Kern,
 Phys. Rev. Lett. {\bf 75}, 677 (1995). 
\bibitem{Schwoebel} R.L. Schwoebel and E.J. Shipsey, J. Appl. Phys.
 {\bf 37},  3682 (1966); G. Ehrlich and F.G. Hudda, J. Chem. Phys. {\bf 44}, 1039 (1966).
\bibitem{Stumpf} R. Stumpf and M. Scheffler, Phys. Rev. Lett. {\bf 72}, 254 (1994).
\bibitem{Perdew} J.P. Perdew et al., Phys. Rev. B {\bf 46}, 6671 (1992).
\bibitem{Stumpf2} R. Stumpf and M. Scheffler, Computer Phys. Commun. {\bf 79},
447 (1994).
\bibitem{Troullier} N. Troullier and J.L. Martins, Solid State Commun. 
 {\bf 74}, 613 (1990); Phys. Rev. B {\bf 43}, 1993 (1991).
\bibitem{Gonze} X. Gonze, R. Stumpf, and M. Scheffler, Phys. Rev. B {\bf 44}, 8503-8513 (1991).
\bibitem{Fuchs} M. Fuchs et al., Computer Phys. Commun., in preparation.
\bibitem{Yu2} B. D. Yu, and M. Scheffler, in preparation.
\bibitem{Meth92b} M.Methfessel, D. Hennig, and M. Scheffler,
 Phys. Rev. B {\bf 46}, 4816 (1992).
\bibitem{Khein} A. Khein, D.J. Singh, and C.J. Umrigar, Phys. Rev. B {\bf 51},
 4105 (1995).
\bibitem{Kley} A. Kley et al., in preparation.
\bibitem{Neugebau92} J. Neugebauer and M. Scheffler, Phys. Rev. B {\bf 46}, 16
067 (1992).
\bibitem{Feibelman} P.J. Feibelman, Phys. Rev. Lett. {\bf 65}, 729 (1990).
\bibitem{Kellogg} G.L. Kellogg and P. Feibelman, Phys. Rev. Lett. 
 {\bf 64}, 3143 (1990). 
\bibitem{Chen} C.L. Chen and T.T. Tsong,  Phys. Rev. Lett. {\bf 64}, 3147 (1990).
\bibitem{Hansen} L. Hansen, P. Stoltze, K.W. Jacobsen and J.K. N\mbox{\o}rskov, 
 Phys. Rev. B {\bf 44}, 6523 (1991).
\bibitem{Boisvert} G. Boisvert, L.J. Lewis, M.J. Puska, and R.M. Nieminen, 
 Phys. Rev. B {\bf 52}, 9078 (1995). 
\bibitem{Liu} C.L. Liu, J.M. Cohen, J.B. Adams, and A.F. Voter, Surf. Sci. 
 {\bf 253}, 334 (1991).
\end{references}
\end{document}